\documentclass[11pt]{article}
\usepackage{amssymb,amsmath,amsfonts}
\usepackage{graphicx}
\usepackage{graphics}
\usepackage{eepic,epsfig}

\makeatletter
\@addtoreset{equation}{section}

\makeatother

\begin{document}

\title{Quasidiscrete spectrum Cherenkov radiation by a charge moving inside
a dielectric waveguide}
\author{A. A. Saharian$^{1,2}$\thanks{%
E-mail: saharian@ysu.am}, S. B. Dabagov$^{3}$, H. F. Khachatryan$^{1}$, L.
Sh. Grigoryan$^{1}$ \vspace{0.3cm} \\
\textit{$^1$Institute of Applied Problems of Physics NAS RA, }\\
\textit{25 Hr. Nersessian Street, 0014 Yerevan, Armenia} \vspace{0.3cm}\\
\textit{$^2$ Institute of Physics, Yerevan State University,}\\
\textit{1 Alex Manoogian Street, 0025 Yerevan, Armenia } \vspace{0.3cm}\\
\textit{$^3$ INFN Laboratori Nazionali di Frascati,}\\
\textit{Via E. Fermi 50, I-00044 Frascati, Italy}}
\maketitle

\begin{abstract}
We investigate the Cherenkov radiation by a charge uniformly moving inside a
dielectric cylindrical channel in a homogeneous medium. The expressions for
the Fourier components of the electric and magnetic fields are derived by
using the electromagnetic field Green tensor. The spectral distribution of
the Cherenkov radiation intensity in the exterior medium is studied for the
general case of frequency dispersion of the interior and exterior dielectric
functions. It is shown that, under certain conditions on the dielectric
permittivities, strong narrow peaks appear in the spectral distribution. The
spectral locations of those peaks are specified and their heights and widths
are estimated analytically on the base of the dispersion equation for the
electromagnetic eigenmodes of the cylinder.
\end{abstract}

\bigskip

\section{Introduction}

For many years after the discovery of Cherenkov radiation (CR, for review
see \cite{Afan04}), it remains an active subject of theoretical and
experimental investigations. The interest in CR is motivated by its
applications as a high-power source of electromagnetic radiation in various
frequency ranges, as well as related to a wide variety of applications in
high-energy physics and materials science. The recent advances in the fields
of metamaterials, nanophysics and photonic crystals open an interesting
possibility to design materials having specified electromagnetic properties
with controllable dispersion relations for effective dielectric permittivity
and magnetic permeability. This provides an effective mechanism to control
the characteristics of CR. Different guiding structures have been used for
the further control of the radiation energy spectral-angular distributions.
They include planar, cylindrical and spherical interfaces between two media
with different electrodynamical properties (references for early
investigations can be found in \cite{Afan04,Bolo62}). More complicated
structures require approximate and numerical methods for the investigation
of the radiation features (see, for example, \cite{Abaj10,Garc04,Tyuk19} and
references therein).

In the present paper we consider the features of CR by a charged particle
moving inside a cylindrical waveguide parallel to its axis. The various
aspect of the energy losses of charged particles interacting with
cylindrical interfaces have been discussed by many authors (see, e.g., \cite%
{Bolo62,Abaj10}, \cite{Zutt86}-\cite{Saha23}, for a more complete list of
references see \cite{Saha20,Saha23}). In particular, these interactions play
an important role in particle accelerators. In addition of standard
conducting and dielectric waveguides, as a realization of the cylindrical
guiding structure we mention here the carbon nanotubes with radii tunable in
relatively wide range.

The paper is organized as follows. In the next section we describe the
problem setup and present the expressions for the Fourier components of the
vector potential and for magnetic and electric fields. Assuming that the
Cherenkov condition in the exterior medium is satisfied, in section \ref%
{sec:Cher} the formula is derived for the spectral density of the radiation,
evaluating the energy flux through a cylindrical surface with large radius.
The features of the radiation intensity are described depending on the
relative permittivity. In section \ref{sec:Conc} the main results of the
paper are summarized.

\section{Electromagnetic fields}

\label{sec:Fields}

We consider a dielectric cylinder with permittivity $\varepsilon _{0}$
embedded in a medium with dielectric permittivity $\varepsilon _{1}$. For
the electromagnetic field source described by the current density $\mathbf{j}%
(x)$ the vector potential $\mathbf{A}(x)$ can be found by using the Green
tensor $G_{il}(x,x^{\prime })$, where $i,l=1,2,3$ and $x$ stands for the
spacetime point $x=(t,\mathbf{R})$. In the Lorentz gauge the corresponding
relation reads
\begin{equation}
A_{i}(x)=-\frac{1}{2\pi ^{2}c}\int dx^{\prime
}\sum_{l=1}^{3}G_{il}(x,x^{\prime })j_{l}(x^{\prime }).  \label{Ai}
\end{equation}%
In the discussion below cylindrical coordinates $(r,\phi ,z)$ with the axis $%
z$ along the cylinder axis will be used. The indices $i,l=1,2,3$ in (\ref{Ai}%
) will correspond to the components $r,\phi ,z$, respectively. For the
source of the radiation we will take the current density%
\begin{equation}
j_{l}(x)=\delta _{l3}qv\delta (r-r_{0})\delta (\phi -\phi _{0})\delta
(z-vt)/r.  \label{jlx}
\end{equation}%
It describes a point charge $q$ moving with constant velocity $v$ parallel
to the axis of the cylinder. Denoting by $r_{c}$ the cylinder radius, we
assume that $r_{0}<r_{c}$.

Having the vector potential $\mathbf{A}(x)$, the magnetic and electric
fields $\mathbf{H}(x)$ and $\mathbf{E}(x)$ are found by the standard
relations. It is convenient to present the discussion in terms of the
partial Fourier components for the fields and Green tensor defined by
\begin{eqnarray}
\mathbf{F}(x) &=&\sum_{n=-\infty }^{\infty }e^{in(\phi -\phi
_{0})}\int_{-\infty }^{\infty }dk_{z}\,e^{ik_{z}(z-vt)}\mathbf{F}%
_{n}(k_{z},r),  \label{Four} \\
G_{il}(x,x^{\prime }) &=&\sum_{n=-\infty }^{\infty }\int_{-\infty }^{\infty
}d\omega \int_{-\infty }^{\infty }dk_{z}\,G_{il,n}(\omega ,k_{z},r,r^{\prime
})e^{in(\phi -\phi ^{\prime })+ik_{z}(z-z^{\prime })-i\omega (t-t^{\prime
})},  \label{GFour}
\end{eqnarray}%
with $\mathbf{F}=\mathbf{A},\mathbf{H},\mathbf{E}$. From (\ref{Ai}), the
simple relation (to simplify the presentation we will omit the arguments of
the Fourier components $F_{nl}=F_{nl}(k_{z},r)$) $%
A_{nl}=-qvG_{l3,n}(vk_{z},k_{z},r,r_{0})/(\pi c)$, is obtained for the
components of the vector potential generated by the source (\ref{jlx}).

By using the expressions for $G_{l3,n}(\omega ,k_{z},r,r^{\prime })$ from
\cite{Grig95} (see also \cite{Saha23}), we get
\begin{equation}
A_{nl}=\frac{qvk_{z}}{2\pi i^{l}c}\frac{J_{n}(\lambda
_{0}r_{0})H_{n}(\lambda _{1}r_{c})}{r_{c}V_{n}^{J}\alpha _{n}(k_{z})}%
\sum_{p=\pm 1}\frac{p^{l-1}}{V_{n+p}^{J}}J_{n+p}(\lambda
_{0}r_{c<})H_{n+p}(\lambda _{1}r_{c>}),  \label{An12}
\end{equation}%
for the components $l=1,2$ and
\begin{eqnarray}
A_{n3} &=&\frac{iqv}{2c}\left[ J_{n}(\lambda _{0}r_{<})H_{n}(\lambda
_{0}r_{>})-\frac{V_{n}^{H}}{V_{n}^{J}}J_{n}(\lambda _{0}r_{0})J_{n}(\lambda
_{0}r)\right] ,\;r<r_{c},\;  \notag \\
A_{n3} &=&-\frac{qv}{\pi c}\frac{J_{n}(\lambda _{0}r_{0})}{r_{c}V_{n}^{J}}%
H_{n}(\lambda _{1}r),\;r>r_{c},  \label{An3}
\end{eqnarray}%
for the axial component. Here, $r_{c<}=\mathrm{min}(r_{c},r)$, $r_{c>}=%
\mathrm{max}(r_{c},r)$, $r_{<}=\mathrm{min}(r_{0},r)$, $r_{>}=\mathrm{max}%
(r_{0},r)$, $J_{n}(y)$ is the Bessel function and $H_{n}(x)=H_{n}^{(1)}(x)$
is the Hankel function of the first kind. Other notations are defined by the
relations
\begin{eqnarray}
V_{n}^{F} &=&F_{n}(\lambda _{0}r_{c})\partial _{r_{c}}H_{n}(\lambda
_{1}r_{c})-H_{n}(\lambda _{1}r_{c})\partial _{r_{c}}F_{n}(\lambda
_{0}r_{c}),\;\lambda _{j}=k_{z}\sqrt{\beta _{j}^{2}-1},  \label{VnF} \\
\alpha _{n}(k_{z}) &=&\frac{\varepsilon _{0}}{\varepsilon _{1}-\varepsilon
_{0}}+\frac{1}{2}\sum_{l=\pm 1}\left[ 1-\frac{\lambda _{1}}{\lambda _{0}}%
\frac{J_{n+l}(\lambda _{0}r_{c})H_{n}(\lambda _{1}r_{c})}{J_{n}(\lambda
_{0}r_{c})H_{n+l}(\lambda _{1}r_{c})}\right] ^{-1},  \label{alfn}
\end{eqnarray}%
for $F=J,H$, $\omega =vk_{z}$ and $\beta _{j}=v\sqrt{\varepsilon _{j}}/c$, $%
j=0,1$. Note that $\lambda _{0}$ and $\lambda _{1}$ are the radial wave
numbers inside and outside the cylinder. It is easily checked that the
components (\ref{An12}) and (\ref{An3}) are continuous on the cylinder
surface $r=r_{c}$. The part of the field coming from the first term in the
square brackets of (\ref{An3}) corresponds to the vector potential generated
by the source (\ref{jlx}) in a homogeneous medium with dielectric
permittivity $\varepsilon _{0}$. The corresponding Fourier component is
expressed as $\mathbf{A}_{n}^{(0)}(k_{z},r)=iq\mathbf{v}J_{n}(\lambda
_{0}r_{<})H_{n}(\lambda _{0}r_{>})/(2c)$.

Given the vector potential, we can find the magnetic and electric fields.
Inside the cylinder, $r<r_{c}$, the Fourier components of the fields are
decomposed into two contributions:%
\begin{equation}
\mathbf{F}_{n}(k_{z},r)=\mathbf{F}_{n}^{(0)}(k_{z},r)+\mathbf{F}%
_{n}^{(1)}(k_{z},r),  \label{Fdec}
\end{equation}%
with $\mathbf{F}=\mathbf{H}$ and $\mathbf{F}=\mathbf{E}$ for the magnetic
and electric fields, respectively. The part $\mathbf{F}_{n}^{(0)}(k_{z},r)$
corresponds to the field of a charged particle moving in an infinite
homogeneous medium with permittivity $\varepsilon _{0}$ and the contribution
$\mathbf{F}_{n}^{(1)}(k_{z},r)$ is induced by the difference of the
permittivity in the region $r>r_{c}$ from $\varepsilon _{0}$. From the
expression for $\mathbf{A}_{n}^{(0)}(k_{z},r)$ given above, omitting the
arguments $(k_{z},r)$, for the first contribution in (\ref{Fdec}) we get
\begin{eqnarray}
H_{nl}^{(0)} &=&-\frac{qv\lambda _{0}}{4c}\sum_{p=\pm 1}\left( \frac{p}{i}%
\right) ^{l-1}\left\{
\begin{array}{ll}
J_{n}(\lambda _{0}r_{0})H_{n+p}(\lambda _{0}r), & r>r_{0} \\
H_{n}(\lambda _{0}r_{0})J_{n+p}(\lambda _{0}r), & r<r_{0}%
\end{array}%
\right. ,  \label{Hnl0} \\
E_{nl}^{(0)} &=&-\frac{q\lambda _{0}}{4\varepsilon _{0}}\sum_{p=\pm 1}\left(
\frac{p}{i}\right) ^{l}\left\{
\begin{array}{ll}
J_{n}(\lambda _{0}r_{0})H_{n+p}(\lambda _{0}r), & r>r_{0} \\
H_{n}(\lambda _{0}r_{0})J_{n+p}(\lambda _{0}r), & r<r_{0}%
\end{array}%
\right. ,  \label{Enl0} \\
H_{n3}^{(0)} &=&0,\;E_{n3}^{(0)}(k_{z},r)=-\frac{qk_{z}}{2\varepsilon _{0}}%
\left( \beta _{0}^{2}-1\right) J_{n}(\lambda _{0}r_{<})H_{n}(\lambda
_{0}r_{>}),  \label{Hn30}
\end{eqnarray}%
with $l=1,2$. For the contributions $\mathbf{F}_{n}^{(1)}$ inside the
cylinder we find%
\begin{eqnarray}
H_{nl}^{(1)} &=&-\frac{qvk_{z}}{2\pi i^{l}c}\sum_{p=\pm
1}p^{l-1}D_{n,p}^{(i)}J_{n+p}(\lambda _{0}r),\;H_{n3}^{(1)}=-\frac{qv\lambda
_{0}}{2\pi c}\sum_{p=\pm 1}pD_{n,p}^{(i)}J_{n}(\lambda _{0}r),  \notag \\
E_{nl}^{(1)} &=&\frac{i^{1-l}qk_{z}}{4\pi \varepsilon _{0}}\sum_{p,p^{\prime
}=\pm 1}p^{l}\left( 1+p^{\prime }\beta _{0}^{2}\right) D_{n,p^{\prime
}p}^{(i)}J_{n+p}(\lambda _{0}r),\;E_{n3}=\frac{iq\lambda _{0}}{2\pi
\varepsilon _{0}}\sum_{p}D_{n,p}^{(i)}J_{n}(\lambda _{0}r),  \label{Hnl}
\end{eqnarray}%
for $l=1,2$. Here we have introduced the notation
\begin{equation}
D_{n,p}^{(i)}=\frac{J_{n}(\lambda _{0}r_{0})}{r_{c}V_{n}^{J}}\left[ pk_{z}%
\frac{H_{n}(\lambda _{1}r_{c})H_{n+p}(\lambda _{1}r_{c})}{\alpha
_{n}(k_{z})V_{n+p}^{J}}-\frac{i\pi \lambda _{0}}{2k_{z}}r_{c}V_{n}^{H}\right]
.  \label{Dni}
\end{equation}

For the Fourier components in the region $r>r_{c}$ one gets%
\begin{eqnarray}
H_{nl} &=&-\frac{qvk_{z}}{2\pi i^{l}c}\sum_{p=\pm
1}p^{l-1}D_{n,p}^{(e)}H_{n+p}(\lambda _{1}r),\;H_{n3}=-\frac{qv\lambda _{1}}{%
2\pi c}\sum_{p}pD_{n,p}^{(e)}H_{n}(\lambda _{1}r),  \notag \\
E_{nl} &=&\frac{i^{1-l}qk_{z}}{4\pi \varepsilon _{1}}\sum_{p,p^{\prime }=\pm
1}p^{l}\left( 1+p^{\prime }\beta _{1}^{2}\right) D_{n,p^{\prime
}p}^{(e)}H_{n+p}(\lambda _{1}r),\;E_{n3}=\frac{iq\lambda _{1}}{2\pi
\varepsilon _{1}}\sum_{p}D_{n,p}^{(e)}H_{n}(\lambda _{1}r),  \label{Hnle}
\end{eqnarray}%
where $l=1,2$, and%
\begin{equation}
D_{n,p}^{(e)}=\frac{J_{n}(\lambda _{0}r_{0})}{r_{c}V_{n}^{J}}\left[ pk_{z}%
\frac{H_{n}(\lambda _{1}r_{c})J_{n+p}(\lambda _{0}r_{c})}{\alpha
_{n}(k_{z})V_{n+p}^{J}}-\frac{\lambda _{1}}{k_{z}}\right] .  \label{Dne}
\end{equation}

\section{Cherenkov radiation in the exterior medium}

\label{sec:Cher}

In this section we consider CR in the region $r>r_{c}$ assuming that the
exterior medium is transparent and the dielectric permittivity $\varepsilon
_{1}$ is real. CR in the exterior medium is present under the condition $%
\beta _{1}^{2}>1$ corresponding to real values of $\lambda _{1}$. For the
corresponding frequency one has $\omega =k_{z}v$ and the radiation
propagates along the Cherenkov angle $\theta =\theta _{\mathrm{Ch}}=\arccos
(1/\beta _{1})$ with respect to the cylinder axis. The radial dependence of
the Fourier components of the fields is expressed in terms of the Hankel
functions $H_{n}(\lambda _{1}r)$ and $H_{n\pm 1}(\lambda _{1}r)$. For $\beta
_{1}^{2}<1$ one has $\lambda _{1}=ik_{z}\sqrt{1-\beta _{1}^{2}}$ and the
radial dependence is described by the Macdonald functions $K_{n}(|\lambda
_{1}|r)$, $K_{n\pm 1}(|\lambda _{1}|r)$ with an exponential decrease of the
Fourier components at large distances from the cylinder surface. This
corresponds to total reflection of the corresponding waves and the cylinder
with dielectric permittivity $\varepsilon _{0}$ is a perfect resonator for
them with discrete eigenvalues for $k_{z}$ and $\omega =k_{z}v$. Those
eigenvalues are roots of the equation $\alpha _{n}(k_{z})=0$. They
correspond to guiding modes of the cylindrical waveguide for $\lambda
_{0}^{2}>0$ and to surface polariton modes for $\lambda _{0}^{2}<0$. The
radiation of surface polaritons in the problem under consideration has been
investigated in \cite{Saha23}.

Assuming that $\beta _{1}^{2}>1$, let us evaluate the energy flux per unit
time through a cylindrical surface of radius $r$ coaxial with the dielectric
cylinder. Denoting by $\mathbf{n}_{r}$ the exterior unit normal to the
surface, the flux is expressed in terms of the Fourier components as
\begin{equation}
I=\pi cr\sum_{n=-\infty }^{\infty }\int_{-\infty }^{\infty }dk_{z}\,\mathbf{n%
}_{r}\cdot \left[ \mathbf{E}_{n}(k_{z},r)\times \mathbf{H}_{n}^{\ast
}(k_{z},r)\right] .  \label{Ifl2}
\end{equation}%
For large values of $r$ we use in the expressions (\ref{Hnle}) the
asymptotic of the Hankel functions for large arguments. The fields are
presented as $\mathbf{F}_{n}(k_{z},r)=\mathbf{F}_{n}^{\mathrm{(r)}%
}(k_{z},r)\left( 1+\mathcal{O}(1/r)\right) $ with $\mathbf{F}=\allowbreak
\mathbf{E},\mathbf{H}$. For the cylindrical components of the radiation
fields $\mathbf{F}_{n}^{\mathrm{(r)}}(k_{z},r)=\mathbf{F}_{n}^{\mathrm{(r)}}$
this gives
\begin{equation}
E_{n1}^{\mathrm{(r)}}=\frac{qk_{z}}{i\pi ^{3/2}\varepsilon _{1}}\frac{%
e^{i\left( \lambda _{1}r-n\pi /2-\pi /4\right) }}{\sqrt{2\lambda _{1}r}}%
\sum_{p=\pm 1}D_{n,p}^{(e)},\;E_{n2}^{\mathrm{(r)}}=-\frac{qk_{z}v^{2}}{\pi
^{3/2}c^{2}}\frac{e^{i\left( \lambda _{1}r-n\pi /2-\pi /4\right) }}{\sqrt{%
2\lambda _{1}r}}\sum_{p=\pm 1}pD_{n,p}^{(e)},  \label{En12rad}
\end{equation}%
and%
\begin{equation}
E_{n3}^{\mathrm{(r)}}=-\sqrt{\beta _{1}^{2}-1}E_{n1}^{\mathrm{(r)}%
},\;H_{n1}^{\mathrm{(r)}}=-\frac{c}{v}E_{n2}^{\mathrm{(r)}},\;H_{n2}^{%
\mathrm{(r)}}=\frac{\varepsilon _{1}v}{c}E_{n1}^{\mathrm{(r)}},\;H_{n3}^{%
\mathrm{(r)}}=\frac{c}{v}\sqrt{\beta _{1}^{2}-1}E_{n2}^{\mathrm{(r)}}.
\label{En3rad}
\end{equation}%
It is easily checked that $\mathbf{E}_{n}^{\mathrm{(r)}}\cdot \mathbf{H}%
_{n}^{\mathrm{(r)}}=0$, $\mathbf{n}_{\mathrm{Ch}}\cdot \mathbf{E}_{n}^{%
\mathrm{(r)}}=0$, and $\mathbf{n}_{\mathrm{Ch}}\cdot \mathbf{H}_{n}^{\mathrm{%
(r)}}=0$ with the unit vector $\mathbf{n}_{\mathrm{Ch}}=(\sqrt{\beta
_{1}^{2}-1},0,1)/\beta _{1}$ determining the direction of the radiation
propagation. These conditions show that the radiation fields describe a
transverse wave with orthogonal electric and magnetic fields. The radiation
fields are decomposed into two polarizations denoted here as $\mathbf{F}%
_{n}^{\mathrm{(r)}\parallel }$ and $\mathbf{F}_{n}^{\mathrm{(r)}\perp }$.
For the parallel polarization the electric field lies in the plane
containing the observation point and the cylinder axis. The corresponding
parts of the fields are expressed as%
\begin{equation}
\mathbf{E}_{n}^{\mathrm{(r)}\parallel }=(1,0,-\sqrt{\beta _{1}^{2}-1}%
)E_{n1}^{\mathrm{(r)}},\;\mathbf{H}_{n}^{\mathrm{(r)}\parallel }=(0,1,0)%
\frac{\varepsilon _{1}v}{c}E_{n1}^{\mathrm{(r)}}.  \label{Epar}
\end{equation}%
For the perpendicular polarization the fields are given by
\begin{equation}
\mathbf{E}_{n}^{\mathrm{(r)}\perp }=(0,1,0)E_{n2}^{\mathrm{(r)}},\;\mathbf{H}%
_{n}^{\mathrm{(r)}\perp }=(-1,0,\sqrt{\beta _{1}^{2}-1})\frac{c}{v}E_{n2}^{%
\mathrm{(r)}}.  \label{Eperp}
\end{equation}

Substituting the radiation fields in (\ref{Ifl2}), the corresponding
spectral density, defined by $I=\int d\omega \,\frac{dI}{d\omega }$ with $%
\omega =k_{z}v$, is presented in the form%
\begin{eqnarray}
\frac{dI}{d\omega } &=&\frac{2q^{2}\omega }{\pi ^{2}vr_{c}^{2}\varepsilon
_{1}}\sideset{}{'}{\sum}_{n=0}^{\infty }\,\left\vert \frac{J_{n}(\lambda
_{0}r_{0})}{V_{n}^{J}}\right\vert ^{2}\left[ \beta _{1}^{2}\left\vert \frac{%
k_{z}H_{n}(\lambda _{1}r_{c})}{\alpha _{n}(k_{z})}\sum_{p}\frac{%
J_{n+p}(\lambda _{0}r_{c})}{V_{n+p}^{J}}\right\vert ^{2}\right.   \notag \\
&&\left. +\left\vert 2\sqrt{\beta _{1}^{2}-1}-\frac{k_{z}H_{n}(\lambda
_{1}r_{c})}{\alpha _{n}(k_{z})}\sum_{p}p\frac{J_{n+p}(\lambda _{0}r_{c})}{%
V_{n+p}^{J}}\right\vert ^{2}\right] ,  \label{dI}
\end{eqnarray}%
where $k_{z}=\omega /v$, $\lambda _{j}=(\omega /v)\sqrt{\beta _{j}^{2}-1}$,
and the prime on the summation sign means that the term $n=0$ should be
taken with coefficient 1/2. The contributions of the first and second terms
in the square brackets of (\ref{dI}) correspond to the spectral densities of
the radiation of waves with perpendicular and parallel polarizations,
respectively. The contribution of the parallel polarization is further
simplified by using the relation%
\begin{equation}
H_{n}(\lambda _{1}r_{c})\sum_{p}p\frac{J_{n+p}(\lambda _{0}r_{c})}{%
V_{n+p}^{J}}=\frac{2}{\lambda _{1}}\left[ \frac{\varepsilon _{1}}{%
\varepsilon _{1}-\varepsilon _{0}}-\alpha _{n}(k_{z})\right] .  \label{rel1}
\end{equation}%
The formula (\ref{dI}) is valid for general case of dispersions $\varepsilon
_{j}=\varepsilon _{j}(\omega )$ for dielectric permittivities and also for
complex valued function $\varepsilon _{0}(\omega )$. The dimensionless
quantity $q^{-2}r_{c}\varepsilon _{1}dI/d\omega $ is a function of $\beta
_{0}$, $\beta _{1}$, and $k_{z}r_{c}=\omega r_{c}/v$. The corresponding
quantity for the radiation in a homogeneous medium with permittivity $%
\varepsilon _{1}$ is obtained in the limit $\varepsilon _{0}\rightarrow
\varepsilon _{1}$. In this limit one has $r_{c}V_{n}^{J}\rightarrow 2i/\pi $%
, $1/\alpha _{n}(k_{z})\propto \varepsilon _{1}/\varepsilon _{0}-1$, and
from (\ref{dI}) we get the standard expression $dI_{1}/d\omega =q^{2}v\omega
\left( 1-\beta _{1}^{-2}\right) /c^{2}$.

In the special case of the motion along the cylinder axis one has $r_{0}=0$
and the only nonzero contribution in (\ref{dI}) comes from the mode $n=0$.
The general formula is reduced to
\begin{equation}
\frac{dI}{d\omega }=\frac{4q^{2}v}{\pi ^{2}c^{2}}\omega \,\left( 1-\frac{1}{%
\beta _{1}^{2}}\right) \left\vert \frac{1}{r_{c}V_{0}^{J}}\right\vert
^{2}\left\vert 1+\frac{1-\varepsilon _{1}/\varepsilon _{0}}{\beta _{1}^{2}-1}%
\left[ 1-\frac{\varepsilon _{1}\lambda _{0}J_{0}(\lambda
_{0}r_{c})H_{1}(\lambda _{1}r_{c})}{\varepsilon _{0}\lambda
_{1}J_{1}(\lambda _{0}r_{c})H_{0}(\lambda _{1}r_{c})}\right]
^{-1}\right\vert ^{2},  \label{dIr0}
\end{equation}%
where $V_{0}^{J}=\lambda _{0}J_{1}(\lambda _{0}r_{c})H_{0}(\lambda
_{1}r_{c})-\lambda _{1}J_{0}(\lambda _{0}r_{c})H_{1}(\lambda _{1}r_{c})$.
For large frequencies, assuming that $\omega r_{c}/v\gg 1$, the arguments of
the cylindrical functions in (\ref{dIr0}) are large and we use the
corresponding asymptotic formulas. To the leading order this gives%
\begin{equation}
\frac{dI}{d\omega }\approx \frac{dI_{1}}{d\omega }\sqrt{\frac{|\beta
_{0}^{2}-1|}{\beta _{1}^{2}-1}}\frac{f_{1}(|\lambda _{0}|r_{c})}{1+\frac{%
\varepsilon _{0}^{2}}{\varepsilon _{1}^{2}}\frac{\beta _{1}^{2}-1}{|\beta
_{0}^{2}-1|}f_{2}(|\lambda _{0}|r_{c})},\;f_{1}(x)=\left\{
\begin{array}{ll}
1+\tan ^{2}\left( x-\pi /4\right) , & \beta _{0}>1 \\
\exp \left( -2x\right) , & \beta _{0}<1%
\end{array}%
\right. ,  \label{freql}
\end{equation}%
with $f_{2}(x)=1$ for $\beta _{0}<1$ and $f_{2}(x)=f_{1}(x)-1$ for $\beta
_{0}>1$. For $\beta _{0}<1$ the spectral density of the radiation intensity
is suppressed by the exponential factor $\exp [-2\left( \omega
r_{c}/c\right) \sqrt{1/\beta ^{2}-\varepsilon _{0}}]$.

For small values of the combination $\omega r_{c}/v$, $\omega r_{c}/v\ll 1$,
additionally assuming that $|\lambda _{j}|r_{c}\ll 1$, from (\ref{dI}) it
can be seen that the contribution of the term with a given $n$ is
proportional to $(\lambda _{0}r_{0})^{2n}$ and the dominant contribution
comes from the term with $n=0$. To the leading order, the radiation
intensity coincides with that in a homogeneous medium with permittivity $%
\varepsilon _{1}$.

\begin{figure}[tbph]
\begin{center}
\begin{tabular}{cc}
\epsfig{figure=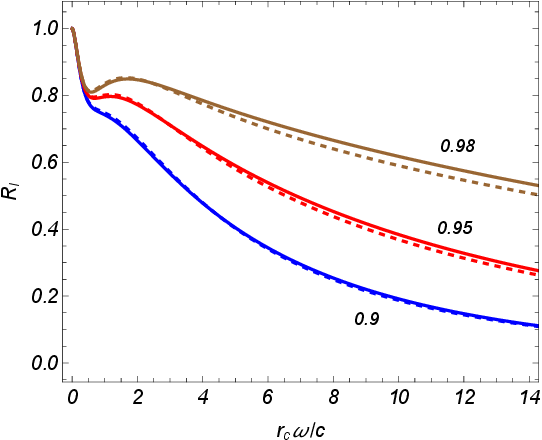,width=7.cm,height=5.5cm} & \quad %
\epsfig{figure=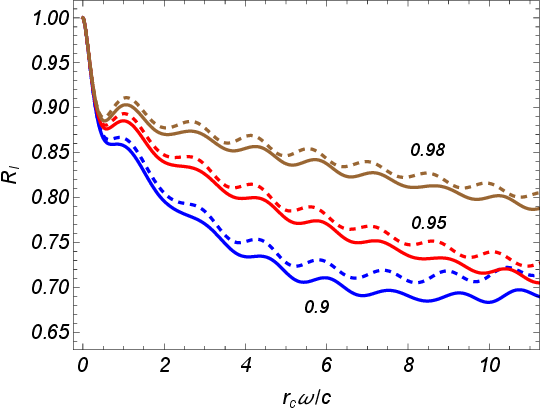,width=7.cm,height=5.5cm}%
\end{tabular}%
\end{center}
\caption{The dependence of the ratio $R_{I}$ on $r_{c}\protect\omega /c$ for
the electron energies $\mathcal{E}_{e}=2\,\mathrm{MeV}$ (solid curves) and $%
\mathcal{E}_{e}=10\,\mathrm{MeV}$ (dashed curves). The graphs are plotted
for $\protect\varepsilon _{1}=3.8$. The left and right panels correspond to $%
\protect\varepsilon _{0}=1$ and $\protect\varepsilon _{0}=2.2$. The numbers
near the curves are the values of $r_{0}/r_{c}$.}
\label{fig1}
\end{figure}

In figure \ref{fig1}, the ratio $R_{I}=(dI/d\omega )/(dI_{1}/d\omega )$ is
plotted versus $r_{c}\omega /c$ for $r_{0}/r_{c}=0.9,0.95,0.98$. For the
electron energy $\mathcal{E}_{e}$ we have taken $\mathcal{E}_{e}=2\,\mathrm{%
MeV}$ (solid curves) and $\mathcal{E}_{e}=10\,\mathrm{MeV}$ (dashed curves).
For dielectric permittivity of the exterior medium the value $\varepsilon
_{1}=3.8$ is taken that corresponds to the dielectric permittivity for fused
quartz in the frequency range $\lesssim 1\,\mathrm{THz}$. The graphs on the
left and right panels are plotted for $\varepsilon _{0}=1$ and $\varepsilon
_{0}=2.2$, respectively. The second value corresponds to dielectric
permittivity for teflon.

In the numerical example presented in figure \ref{fig1} we have taken real
dielectric permittivities obeying the condition $\varepsilon
_{0}<\varepsilon _{1}$. In the corresponding spectral range the intensity is
smaller than the one for CR in a homogeneous medium with permittivity $%
\varepsilon _{1}$. For the case corresponding to the left panel (the charge
moves in a vacuum cylindrical hole), the CR is formed in the exterior
medium. For the example on the right panel the Cherenkov condition is obeyed
in both regions $r<r_{c}$ and $r>r_{c}$ and the spectral distribution of the
CR intensity exhibits oscillations. These oscillations originate from the
interference of CR formed inside the cylinder. For relativistic velocities
the dependence on the energy of the particles is relatively week. Note that
for the values of the parameters corresponding to the right panel of figure %
\ref{fig1} we have $(dI_{0}/d\omega )/(dI_{1}/d\omega )\approx 0.74$ for $%
\mathcal{E}_{e}=10\,\mathrm{MeV}$ ($\approx 0.72$ for the energy $2\,\mathrm{%
MeV}$), where $dI_{0}/d\omega $ is the spectral density for CR in a
homogeneous medium with dielectric permittivity $\varepsilon _{0}$. This
value is not far from the value of $R_{I}$ for large $\omega r_{c}/c$.

\begin{figure}[tbph]
\begin{center}
\begin{tabular}{cc}
\epsfig{figure=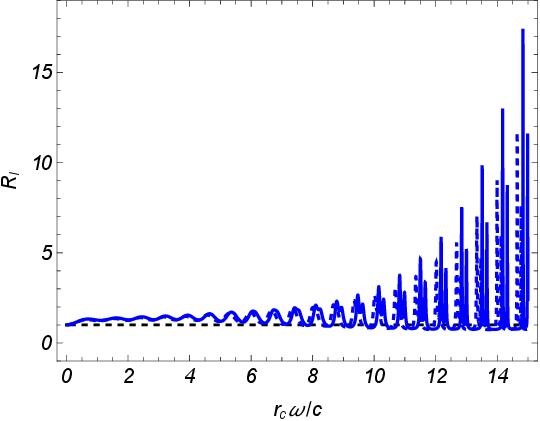,width=7.cm,height=5.5cm} & \quad %
\epsfig{figure=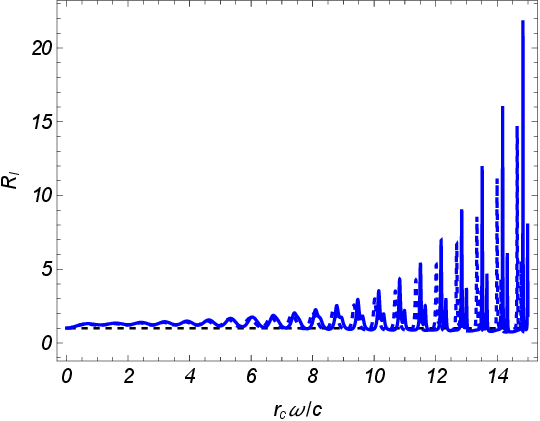,width=7.cm,height=5.5cm}%
\end{tabular}%
\end{center}
\caption{The same as in figure \protect\ref{fig1} for $\protect\varepsilon %
_{0}=3.8$, $\protect\varepsilon _{1}=2.2$. The left and right panels are
plotted for $r_{0}/r_{c}=0.9$ and $r_{0}/r_{c}=0.95$.}
\label{fig2}
\end{figure}

Figure \ref{fig2} presents the spectral distribution of the ratio $R_{I}$
for $\varepsilon _{0}=3.8$ and $\varepsilon _{1}=2.2$. As before, the solid
and dashed curves correspond to the particle energies $\mathcal{E}_{e}=2\,%
\mathrm{MeV}$ and $\mathcal{E}_{e}=10\,\mathrm{MeV}$, respectively (note
that the values of $\varepsilon _{0}$ and $\varepsilon _{1}$ are transposed
for the right panel in figure \ref{fig1} and for figure \ref{fig2}). We have
taken $r_{0}/r_{c}=0.9$ for the left panel and $r_{0}/r_{c}=0.95$ for the
right one. As seen from figure \ref{fig2}, in this case with $\varepsilon
_{0}>\varepsilon _{1}$, strong narrow peaks appear in the spectral
distribution of CR in the exterior medium for large values of $r_{c}\omega
/c $.

This feature of the radiation intensity distribution in the cylindrical
setup under consideration has also been observed for a charge moving outside
a cylinder \cite{Saha20} and in the spectral-angular distribution of the
radiation by a charge circulating around or inside a cylinder \cite%
{Saha05,Saha09,Kota13}. The analytic explanation for the presence of the
strong peaks seen in figure \ref{fig2} is similar to that described in \cite%
{Saha20} and we shortly outline it assuming that both the media inside and
outside the cylinder are transparent.

As noted above, the equation $\alpha _{n}(k_{z})=0$ determines the
eigenmodes of the dielectric cylinder. This equation has solutions under the
condition $\lambda _{1}^{2}<0$ with exponential suppression of the
electromagnetic fields in the region $r>r_{c}$ at distances from the
cylinder surface larger than the radiation wavelength. Two types of the
eigenmodes correspond to guiding modes with $\lambda _{0}^{2}>0$ (CR
confined inside the cylinder) and to surface polaritons with $\lambda
_{0}^{2}<0$. The modes of the second type are present under the condition $%
\varepsilon _{0}\varepsilon _{1}<0$ and the corresponding radiation
intensity has been investigated in \cite{Saha23} assuming that $\varepsilon
_{1}<0<\varepsilon _{0}$. For $\lambda _{1}^{2}>0$ the function $\alpha
_{n}(k_{z})$ appearing in (\ref{dI}) is a complex function and has no zeros
for real $k_{z}$. However, that function can be exponentially small for
large values of $n$. The mathematical reason for that possibility is based
on the fact that for large $n$ the ratio $|J_{n}(ny_{1})/Y_{n}(ny)|$, with
the Neumann function $Y_{n}(u)$, is exponentially small for $0<y<1$.
Assuming that $\lambda _{1}r_{c}<n$, for large $n$ we can expand the
function $\alpha _{n}(k_{z})$ over the small ratio $J_{n+l}(\lambda
_{0}r_{c})/Y_{n+l}(\lambda _{1}r_{c})$ of the Bessel and Neumann functions: $%
\alpha _{n}(k_{z})=\alpha _{n}^{(Y)}(k_{z})+\mathcal{O}(e^{-2n\zeta (\lambda
_{1}r_{c}/n)})$, where%
\begin{equation}
\alpha _{n}^{(Y)}(k_{z})=\frac{\varepsilon _{0}}{\varepsilon
_{1}-\varepsilon _{0}}+\frac{1}{2}\sum_{l=\pm 1}\left[ 1-\frac{\lambda _{1}}{%
\lambda _{0}}\frac{J_{n+l}(\lambda _{0}r_{c})Y_{n}(\lambda _{1}r_{c})}{%
J_{n}(\lambda _{0}r_{c})Y_{n+l}(\lambda _{1}r_{c})}\right] ^{-1},
\label{alfY}
\end{equation}%
and $\zeta (y)=\ln [(1+\sqrt{1-y^{2}})/y]-\sqrt{1-y^{2}}$. Now the function $%
\alpha _{n}^{(Y)}(k_{z})$ is real and at its possible zeros the function $%
1/\alpha _{n}(k_{z})$ in (\ref{dI}) is of the order $e^{2n\zeta (\lambda
_{1}r_{c}/n)}$. Let us denote by $k_{z}r_{c}=u_{n,s}$ the roots of the
equation $\alpha _{n}^{(Y)}(k_{z})=0$ with respect to $k_{z}r_{c}$, where $%
s=1,2,\ldots $ enumerates the roots for a given $n$. Though these roots are
not the eigenmodes of the dielectric cylinder, they obey the eigenmode
equation $\alpha _{n}(k_{z})=0$ with exponential accuracy. In this sense, we
refer to those roots as quasimodes of the cylinder. We have checked
numerically that the locations of the strong peaks in figure \ref{fig2}
coincide with $u_{n,s}v/c$ with high accuracy. Note that the locations of
the peaks do not depend on $r_{0}$. The locations are also the same for the
parallel and perpendicular polarizations of the radiation field.

As it has been shown in \cite{Saha20}, for $\lambda _{0}^{2}>0$ the
necessary condition for the existence of roots for the equation $\alpha
_{n}^{(Y)}(k_{z})=0$ is reduced to $\lambda _{0}r_{c}>n$ which implies that $%
\varepsilon _{0}>\varepsilon _{1}$. In the case $\lambda _{0}^{2}<0$, for
the corresponding necessary condition one gets $\varepsilon
_{0}<-\varepsilon _{1}$. By using the asymptotic expressions of the cylinder
functions we can show that for $\lambda _{0}^{2}>0$ the height of the peak
for a given $n$ is proportional to the factor $e^{2n\zeta (\lambda
_{1}r_{c}/n)}$ for $\lambda _{0}r_{0}>n$. For $\lambda _{0}r_{0}<n$ the
contribution of the term with a given $n$ to $dI/d\omega $ is proportional
to the factor $e^{2n\left[ \zeta (\lambda _{1}r_{c}/n)-\zeta (\lambda
_{0}r_{0}/n\right] }$. By taking into account that the function $\zeta (y)$
is monotonically decreasing in the region $0<y\leq 1$, we conclude that in
order to have a peak in the region $\lambda _{0}r_{0}<n$ the condition $%
\lambda _{0}>\lambda _{1}r_{c}/r_{0}$ is required. For a given velocity,
this condition gives the lower limit for the ratio $r_{0}/r_{c}$: $%
r_{0}/r_{c}>\sqrt{\beta _{1}^{2}-1}/\sqrt{\beta _{0}^{2}-1}$. The peaks
appear for large $n$ and we could expect the presence of the lower limit by
taking into account that for small values of the ratio $r_{0}/r_{c}$ the
corresponding contribution to the radiation intensity contains the factor $%
(r_{0}/r_{c})^{2n}$ coming from the function $J_{n}(\lambda _{0}r_{0})$ in (%
\ref{dI}). The widths of the spectral peaks are obtained expanding the
function $\alpha _{n}(k_{z})$ near the roots $u_{n,s}$ and they are
estimated as $\Delta \omega /\omega \propto e^{-2n\zeta (\lambda
_{1}r_{c}/n)}$ (see also \cite{Saha20}).

It follows from the above explanation that the narrow peaks in figure \ref%
{fig2} correspond to large values of $n$. The numerical analysis shows that
for a given $n$ we have two peaks. The peak with smaller value of $%
r_{c}\omega /c$ is higher. For example, in the case of $\mathcal{E}_{e}=2\,%
\mathrm{MeV}$, $r_{0}/r_{c}=0.95$ (corresponding to the right panel in
figure \ref{fig2}) and for $n=\left\{ 17,18,19,20,21,22\right\} $ one has $%
R_{I}=\left\{ 6.2,8.3,11.2,15.3,21.1,29.1\right\} $ and $R_{I}=\left\{
2.9,3.9,5.3,7.3,10.2,14.2\right\} $ for the peaks with smaller and larger
values of $r_{c}\omega /c$, respectively. The corresponding spectral
locations are given by $r_{c}\omega /c=\left\{
12.18,12.85,13.51,14.18,14.84,15.5\right\} $ and $r_{c}\omega /c=\left\{
13.0,13.67,14.33,14.99,15.65,16.3\right\} $. As seen from these data, the
peaks corresponding to the modes with $n$ and $n+1$ are approximately
equidistant. The relative deviations from equidistance are of the order of $%
1/n$. These features can also be seen analytically using in (\ref{alfY}) the
asymptotic formulas of the Bessel and Neumann functions for large values of
the order. From those formulas it follows that%
\begin{eqnarray}
\frac{J_{n+l}(ny_{0})}{J_{n}(ny_{0})} &\approx &\frac{1}{y_{0}}\left\{ 1+l%
\sqrt{y_{0}^{2}-1}\tan \{n[\sqrt{y_{0}^{2}-1}-\arccos (1/y_{0})]-\pi
/4\}\right\} ,  \notag \\
\frac{Y_{n+l}(ny_{1})}{Y_{n}(ny_{1})} &\approx &\frac{1}{y_{1}}\left( 1+l%
\sqrt{1-y_{1}^{2}}\right) ,  \label{JYas}
\end{eqnarray}%
for $y_{1}<1<y_{0}$. For the functions in (\ref{alfY}) we have $%
y_{j}=(r_{c}\omega /cn)\sqrt{\varepsilon _{j}-1/\beta ^{2}}$. With the
asymptotics (\ref{JYas}), the equation $\alpha _{n}^{(Y)}(k_{z})=0$ is
reduced to a quadratic equation with respect to the tan function in (\ref%
{JYas}). The peaks for a given $n$ correspond to two roots of that equation.
The function $\zeta (y)$ is monotonically decreasing and its value $\zeta
(\lambda _{1}r_{c}/n)$ for the peak with larger $r_{c}\omega /c$ is smaller.
In the numerical example we have considered ($\mathcal{E}_{e}=2\,\mathrm{MeV}
$, $r_{0}/r_{c}=0.95$), for the peaks one has $\lambda _{0}r_{0}>n$ and the
corresponding heights are proportional to $e^{2n\zeta (\lambda _{1}r_{c}/n)}$%
. For a given $n$, this explains the smaller heights for peaks with larger
values of $r_{c}\omega /c$. The peaks appear in the spectral range
determined by
\begin{equation}
n(\varepsilon _{0}-1/\beta ^{2})^{-1/2}<r_{c}\omega /c<n(\varepsilon
_{1}-1/\beta ^{2})^{-1/2}.  \label{peak}
\end{equation}%
For $\varepsilon _{0}=3.8$ and $\varepsilon _{1}=2.2$ this corresponds to
the regions $r_{c}\omega /c\in (0.61n,0.94n)$ and $r_{c}\omega /c\in
(0.6n,0.91n)$ for the energies $\mathcal{E}_{e}=2\,\mathrm{MeV}$ and $%
\mathcal{E}_{e}=10\,\mathrm{MeV}$, respectively. Note that in the region $%
0<y<1$ the large $n$ asymptotic for the Bessel function has the form $%
J_{n}(ny)\approx (1-y^{2})^{-1/4}e^{-n\zeta (y)}/\sqrt{2\pi n}$. Introducing
$r_{b0}=n/\lambda _{0}$ and by taking into account that the radial
dependence of the fields inside the cylinder is expressed in terms of the
Bessel function with the argument $\lambda _{0}r$, we see that in the region
$r<r_{b0}$ and for large $n$ the fields are suppressed by the factor $%
e^{-n\zeta (r/r_{b0})}$. In the asymptotic of the function $J_{n}(ny)$ for $%
y>1$ the exponent $e^{-n\zeta (y)}$ is replaced by $2\cos (g(y))$, where the
function $g(y)$ is given by the argument of the tan function in (\ref{JYas})
with the replacement $y_{0}\rightarrow y$. This cos function determines the
radial dependence of the fields in the region $r_{b0}<r<r_{c}$. In a similar
way, we can introduce $r_{b1}=n/\lambda _{1}>r_{c}$, which separates two
regions of the radial coordinate, $r_{c}<r<r_{b1}$ and $r>r_{b1}$, with
qualitatively different behavior of the fields as functions of the radial
coordinate. The factor determining the heights of the peaks is expressed as $%
e^{2n\zeta (r_{c}/r_{b1})}$. Note that the necessary condition on the ratio $%
r_{0}/r_{c}$ for the appearance of the peaks is written as $%
r_{0}/r_{c}>r_{b0}/r_{b1}$.

The estimates for the characteristics of the spectral peaks were given under
the assumption of real $\varepsilon _{0}$. For complex dielectric
permittivity, in the expansion of the function $\alpha _{n}(k_{z})$ near the
roots $u_{n,s}$ additional terms appear proportional to the ratio $%
\varepsilon _{0}^{\prime \prime }/\varepsilon _{0}^{\prime }$, where $%
\varepsilon _{0}^{\prime }$ and $\varepsilon _{0}^{\prime \prime }$ are the
real and imaginary parts of $\varepsilon _{0}$. The estimates given above
are valid in the range $|\varepsilon _{0}^{\prime \prime }/\varepsilon
_{0}^{\prime }|\ll e^{-2n\zeta (\lambda _{1}r_{c}/n)}$. For the case $%
|\varepsilon _{0}^{\prime \prime }/\varepsilon _{0}^{\prime }|>e^{-2n\zeta
(\lambda _{1}r_{c}/n)}$, the characteristics of the peaks in the spectral
distribution are determined by the ratio $\varepsilon _{0}^{\prime \prime
}/\varepsilon _{0}^{\prime }$. In addition to the decrease of the heights,
the inclusion of the imaginary part of dielectric permittivity leads to
broadening of the peaks. The influence of several other factors, such as the
finite length of the radiator and multiple scattering, has been discussed in
\cite{Saha20} for $r_{0}>r_{c}$. In particular, similar to the case of CR in
a plate of thickness $L$, we expect that the finite length of the radiator
will lead to the angular distribution of the radiation intensity near the
Cherenkov angle in the form of the factor $\sin ^{2}[L\omega (1-\beta
_{1}\cos \theta )/(2v)]/(1-\beta _{1}\cos \theta )^{2}$. In general, we
expect that the results given above will approximate the properties of the
CR in finite thickness radiators for the length $L$ much larger than the
cylinder radius and the radiation wavelength. The finite length appears as
an additional factor that will restrict the exponential increase of the
heights of the spectral peaks of CR for large frequencies. The corresponding
problem is complicated and the detailed influence on the parameters of the
peaks requires a separate investigation.

\section{Conclusion}

\label{sec:Conc}

We have described the features of radiation by a point charge moving inside
a dielectric cylinder immersed in a homogeneous medium. The charge moves
parallel to the cylinder axis at a constant velocity. Depending on the
dielectric functions of the interior and exterior media and on the spectral
range, three different types of polarization radiation can be emitted. They
correspond to CR propagating in the exterior medium, CR confined inside the
cylinder and surface polaritons confined near the cylinder surface. The
radiation of the surface polaritons has been recently considered in \cite%
{Saha23} and here we were mainly concerned with CR in the exterior medium.
The Fourier coefficients for the vector potential and for the electric and
magnetic fields in both exterior and interior regions are determined by
using the Green tensor from \cite{Grig95}. The spectral density of the
intensity for CR in the exterior medium is given by the expression (\ref{dI}%
). The dependence of the radiation intensity on the distance of the charge
trajectory from the axis of the cylinder enters through the function $%
|J_{n}(\lambda _{0}r_{0})|^{2}$ for a given $n$. In the special case of
motion along the cylinder axis the modes with $n=0$ contribute only and the
general formula is further simplified to (\ref{dIr0}).

The spectral distribution of the radiation intensity for CR in the exterior
medium essentially depends on the relation between the exterior and interior
dielectric permittivities. The numerical results have been displayed for the
ratio of spectral distributions in the presence of the cylinder and the
corresponding quantity in a homogeneous medium with permittivity $%
\varepsilon _{1}$. For the case $\varepsilon _{0}<\varepsilon _{1}$, that
ratio is plotted in figure \ref{fig1} and in the corresponding spectral
range the presence of the cylinder leads to the decrease of the CR intensity
compared to the radiation in a homogeneous medium. The situation can be
essentially different in the case $\varepsilon _{0}>\varepsilon _{1}$. As it
is demonstrated by the example in figure \ref{fig2}, strong narrow peaks may
appear in the spectral distribution of the radiation intensity. We have
analytically argued the appearance of the peaks and estimated the
corresponding heights and widths. The peaks come from the contribution of
the modes with large $n$ and their locations coincide with the zeros of the
function (\ref{alfY}) with high accuracy.

We emphasize that the formula (\ref{dI}) is valid for the general case of
the frequency dependence of the dielectric functions $\varepsilon
_{0}(\omega )$ and $\varepsilon _{1}(\omega )$ and also for a complex
function $\varepsilon _{0}(\omega )$. It will be interesting to discuss the
features described above for specific dispersion laws and also in the
spectral ranges where $\lambda _{0}^{2}<0$. Another direction of
investigations corresponds to the study of CR confined inside the cylinder.
That radiation is emitted on the guiding modes of the cylinder being the
solutions of the equation $\alpha _{n}(k_{z})=0$ in the spectral range
corresponding to $\lambda _{1}^{2}<0<\lambda _{0}^{2}$. These investigations
will be presented elsewhere.

\section*{Acknowledgement}

The work was supported by the Higher Education and Science Committee of RA,
in the frames of the projects 21AG-1C047 (A.A.S.) and 21AG-1C069 (L.Sh.G.
and H.F.K.).

\end{document}